\documentclass{article}

\usepackage{a4wide}

\def\ket#1{\mbox{$| #1 \rangle$}}

\def\beforethm{\vspace*{3mm}}       %
\def\afterthm{\vspace*{3mm}}        %

\title{\textbf{Multiparty Quantum Communication Complexity}}

\author{\emph{Harry Buhrman}\thanks{Quantum Computing and Advanced Systems
Research, Centrum voor Wiskunde en Informatica, Kruislaan 413, 
P.O.~Box~94079, 1090~GB~Amsterdam, The~Netherlands. 
Electronic address: \texttt{Harry.Buhrman@cwi.nl}}
\and \emph{Wim van Dam}\thanks{C.W.I. Amsterdam and 
Centre for Quantum Computation, 
Clarendon Laboratory, Department of Physics,
University of Oxford, Parks Road, Oxford OX1~3PU, United Kingdom.
Electronic address: \texttt{wimvdam@qubit.org}}
\and \emph{Peter H{\o}yer}\thanks{BRICS, Department of Computer Science, 
University of Aarhus, Ny~Munkegade, Bldg.~540, DK-8000 Aarhus~C, Denmark.
Electronic address: \texttt{hoyer@brics.dk}}
\and \emph{Alain Tapp}\thanks{D\'epartement d'Informatique et de 
recherche op\'erationnelle, Universit{\'e} de Montr\'eal,
C.P.~6128, succ.~Centre-Ville, Montr\'eal (Qu{\'e}bec), Canada H3C~3J7.
Electronic address: \texttt{tappa@iro.umontreal.ca}}}

\date{March~31, 1999}

\begin{document}
\maketitle

\begin{abstract}
Quantum entanglement cannot be used to achieve direct communication
between remote parties, but it can reduce the communication needed 
for some problems.
Let each of $k$ parties hold some partial input data to some fixed 
$k$-variable function~$f$.  The communication complexity of $f$ is the
minimum number of classical bits required to be broadcasted for every 
party to know the value of $f$ on their inputs.

We construct a function $G$ such that for the one-round communication
model and three parties, $G$ can be computed with $n+1$ bits of
communication when the parties share prior entanglement. 
We~then show that without entangled particles, the
one-round communication complexity of~$G$ is $(3/2)n +1$.
Next we generalize this function to a function~$F$. 
We show that if the parties share prior quantum entanglement, 
then the communication complexity of $F$ is exactly~$k$.
We also show that if no entangled particles are provided, 
then the communication complexity of $F$ is roughly $k \log_2 k$.  

These two results prove for the first time communication complexity 
separations better than a constant number of bits.
\end{abstract}


\section{Introduction}
Suppose each of $k$ parties holds some data that is unknown to the
others, and they want to evaluate some fixed $k$-variable function on
those
data.  If the function is non-trivial, then this cannot be done unless
the
parties communicate.

In~\cite{CB97}, Cleve and Buhrman raised the question whether 
or not less communication is needed if the parties possess entangled
particles. 
They demonstrated that, for a specific problem, prior quantum
entanglement 
decreases the need for communication by 1~bit from 3 to 2~bits.
A~1-bit saving was also obtained by Buhrman, Cleve, and van~Dam 
in~\cite{BCD97} for another problem
where each party initially holds a 2-bit input-string.
In~both of these problems, there are 3 parties ($k=3$).
They left open the important question if a separation larger than
1~bit is possible.  In particular, is a separation in an asymptotic
setting possible?  In this article we show that this is indeed
the case.

Let $f$ be a $k$-variable Boolean function whose inputs 
are $n$-bit binary strings
(that is, $f: X^k \rightarrow \{0,1\}$ where $X = \{0,1\}^n$).
There are $k$ parties, denoted $P_1,\ldots,P_k$,
where party $P_i$ holds input data $x_i$ ($i=1,\ldots,k$).  
Initially, party $P_i$ only knows $x_i$,
so, to evaluate $f$, the parties have to communicate among each other.
The communication is done by broadcasting classical bits, where, 
each time, a party broadcasts one bit to everybody,
on the total cost of one bit of communication.

We are interested in determining the minimum number of bits required 
to be broadcasted in the worst-case
for every party to know the value of~$f$. 
This number is called the {\em communication complexity
of $f$\/} and is denoted $C(f,k,n)$. 
We want to compare this number with $Q(f,k,n)$,
the communication complexity of $f$ {\em with\/} prior quantum
entanglement.  
That is, the situation where we allow the parties to
share a set of entangled particles before they learn their 
inputs~\cite{CB97,BCD97}.

For example, with this terminology, the separation obtained 
in~\cite{BCD97} reads: there exists a 3-variable ($k=3$) 
Boolean function $g$ whose inputs are 2-bit strings ($n=2$), 
and for which $C(g,3,2) = 3$, but $Q(g,3,2) = 2$.
For some functions, no separation at all is possible.
For example, Cleve {\em et~al.}~\cite{CDNT98}
showed that prior quantum entanglement does not help
in computing the so-called inner product function.

References~\cite{CB97,BCD97} left open the very interesting 
question if a separation in an asymptotic setting is possible.  
This question can be phrased more formal as:  
Does there exists a function~$f$ for which 
$C(f,k,n)$ grows in $k$ or $n$, and for which the ratio between
$C(f,k,n)$
and $Q(f,k,n)$ is bounded from below by some constant larger than~1?

In~this paper, we first study the case where the number of parties
is three ($k=3$).  In~this setting we consider the {\em one-round\/}
communication model where each party is allowed to
communicate at most once.
We construct a Boolean function~$G$ 
for which $C(G,3,n) = (3/2)n +1$ whereas $Q(G,3,n) = n+1$. 
This gives a separation by a factor of $3/2$ in terms of
the number of bits hold by each of the three parties.

Next we relax the requirement that only one round of
communication is allowed and consider an arbitrary number of
parties.  To this end we generalize the communication function
$G$ to~$F$.
We demonstrate that
the communication complexity of~$F$ with prior quantum
entanglement is exactly $k$ [that is, $Q(F,k,n)=k$],
but that, if $n \geq \log_2 k$, then 
without quantum entanglement it is roughly $k \log_2 k$ 
[that is, $C(F,k,n) \approx k \log_2 k$].
We prove this by giving upper and lower bounds in both cases.  
This implies a separation by a logarithmic factor in~$k$,
the number of parties.

This paper thus presents a function with a separation 
by a constant factor in terms of the number of bits, 
and a function with a separation by a logarithmic factor
in terms of the number of parties.
Very recently, much more impressive separations have been 
obtained in terms of the number of bits.
Buhrman, Cleve, and Wigderson~\cite{BCW98}, 
Ambainis {\em et~al.}~\cite{ASTVW98}, and
Raz~\cite{Raz99} have all found two-party computational 
problems for which an exponential separation holds.


\section{The Modulo-4 Sum Problem}
In~this section, 
we~fix the number of parties to three ($k=3$).
As~common, we name the parties Alice, Bob, and Carol.

In~\cite{BCD97}, Buhrman, Cleve, and van~Dam considered the
{\em Modulo-4 Sum Problem\/} defined as follows.
Alice, Bob, and Carol receive $x$, $y$, and~$z$, respectively, 
where $x,y,z \in U = \{0,1,2,3\}$, and they are promised that
\begin{equation}\label{eq:singlepromise}
(x+y+z) \textrm{ mod 2} = 0.
\end{equation}
The common goal is for every party to learn the value 
of the function 
\begin{equation}\label{eq:modulofoursum}
f(x,y,z) = \frac12 \Big[(x+y+z) \textrm{ mod 4}\Big].
\end{equation}
We say that $(x,y,z) \in U \times U \times U$ 
is a {\em valid\/} input if Eq.~\ref{eq:singlepromise} holds.
The function $f : U \times U \times U \rightarrow \{0,1\}$
can be viewed as computing the second-least
significant bit in the sum of $x$, $y$, and~$z$.
Note that for all inputs $y,z \in U$ to Bob and Carol, 
there exists a unique input $x \in U$ for Alice such that 
$(x,y,z)$ is a valid input and $f(x,y,z) =1$.

For every integer $m \geq 1$, we generalize~$f$ to
$G_m : U^m \times U^m \times U^m \rightarrow \{0,1\}$ 
by setting
\[ G_m({\mathbf x},{\mathbf y},{\mathbf z}) = 1  
\quad   
   \textnormal{ if and only if }
\quad
\textnormal{ for all $1 \leq i \leq m$
   we have $f(x_i,y_i,z_i) = 1$},
\]
where ${\mathbf x} = (x_1,\ldots,x_m)$, 
${\mathbf y} = (y_1,\ldots,y_m)$, and
${\mathbf z} = (z_1,\ldots,z_m)$,
and with the condition that, 
\begin{equation}\label{eq:validation}
(x_i+y_i+z_i) \textrm{ mod 2} = 0 \qquad \textrm{($1 \leq i \leq m$).} 
\end{equation}
Thus, we give Alice, Bob, and Carol $m$ valid instances of~$f$,
all at the same time, and ask if they all evaluate to~1.
Again, we say that $({\mathbf x},{\mathbf y},{\mathbf z})$ is
a {\em valid\/} input if Eq.~\ref{eq:validation} holds.

Buhrman {\em et~al.}~\cite{BCD97} showed that
{\em with\/} prior entanglement, function~$f$
can be solved with one-round communication using 3 bits.
In~their protocol, Bob and Carol each broadcast one bit,
where after Alice is capable of computing the value of~$f$ and
then broadcasting the resulting bit.
(See Section~\ref{subsec:withentanglement} for a direct 
generalization of their protocol.)
Their protocol therefore immediately yields 
a $2m+1$ bits protocol for~$G_m$.

\beforethm 
{\noindent \bf Theorem~1\quad}{\sl
With prior quantum entanglement $G_m$ can be solved with one-round 
communication using $2m+1$ bits.}
\afterthm

In~Subsection~\ref{subsec:oneroundlowerbound} below, 
we prove the following 
lower bound for the case that we do not allow quantum entanglement.

\beforethm
{\noindent \bf Theorem~2\quad}{\sl
Without quantum entanglement,
there is no one-round protocol for $G_m$ that uses 
less than $3m+1$ bits of communication.}
\afterthm

For one-round protocols we thus archive a separation of $2m+1$ 
bits against $3m+1$ bits.
We do not know the classical communication complexity 
of computing~$G_m$ without any restriction on the number of rounds.

\subsection{Classical Upper Bound}
The lower bound in Theorem~2 is tight
as there is a straightforward one-round protocol 
that computes~$G_m$ with $3m+1$ bits of communication.  
It~is instructive for understanding the proof of our lower bound,
first to understand that protocol.

Consider an input ${\mathbf x} \in U^m$ to Alice.
We~can think of~${\mathbf x} = (x_1,\ldots,x_m)$ 
as consisting of two parts, 
the high bits and the low bits.
That is, we identify $\mathbf x$ with the pair
$({\mathbf x}_{\mathrm{high}},{\mathbf x}_{\mathrm{low}})$,
where the $i$-th coordinate 
in ${\mathbf x}_{\mathrm{high}} \in \{0,1\}^m$
is $(x_i \textrm{ div }2)$,
and where the $i$-th coordinate 
in ${\mathbf x}_{\mathrm{low}} \in \{0,1\}^m$
is $(x_i \textrm{ mod }2)$.
We think of Bob's input~${\mathbf y} = (y_1,\ldots,y_m)$ 
and Carol's input~${\mathbf z} = (z_1,\ldots,z_m)$ 
in a similar manner.

The $3m+1$ one-round protocol works as follows:
First Bob broadcasts all $2m$ bits of his input
$({\mathbf y}_{\mathrm{high}},{\mathbf y}_{\mathrm{low}})$.
Then Carol broadcasts the $m$ high bits
${\mathbf z}_{\mathrm{high}}$ of her input.
Now Alice is capable of computing the value of~$f$ on
all $m$ instances, that is, she can compute
$f(x_i,y_i,z_i)$ for all $1 \leq i \leq m$.
Due to the promise that $(x_i + y_i + z_i) \textrm{ mod~2} = 0$,
she does not need the low bits 
${\mathbf z}_{\mathrm{low}}$ of Carol's input.
Finally Alice checks if $f(x_i,y_i,z_i)=1$ for all $1 \leq i \leq m$. 
If~so, $G_m({\mathbf x},{\mathbf y},{\mathbf z})=1$ and 
Alice therefore broadcasts~1, otherwise she broadcasts~0.

Intuitively, Alice has to have all of Bob's $m$ high bits,
all of Carol's $m$ high bits, 
but just $m$ of the $2m$ low bits.
Hence, {\em intuitively}, if there exists a protocol for~$G_m$
in which Bob broadcasts $s_B$ bits and Carol broadcasts $s_C$ bits,
then $s_B$ should be at least~$m$, $s_C$ at least~$m$, and
$s_B + s_C$ at least~$3m$.
It~is the result of the following subsection that this intuition
is valid.

\subsection{Classical Lower Bound}
\label{subsec:oneroundlowerbound}
We~now prove our lower bound stated in Theorem~2.  
Since we only consider one-round protocols, 
we can without loss of generality assume that any protocol 
computing~$G_m$ is made up of the following three parts:
\begin{enumerate}
\item Bob (knowing only his input~${\mathbf y}$) broadcasts 
      the message $\sigma_B = \sigma_B({\mathbf y})$.
\item Carol (knowing her input~$\mathbf z$ and Bob's 
      message $\sigma_B$)
      broadcasts the message 
      $\sigma_C = \sigma_C({\mathbf z},\sigma_B)$.
\item Alice (knowing ${\mathbf x}$, $\sigma_B$, and $\sigma_C$)
      computes the answer 
      $\sigma_A( {\mathbf x}, \sigma_B,\sigma_C) \in \{0,1\}$
      which she then broadcasts to Bob and Carol.
      Since this protocol computes $G_m$, we can without loss of
      generality assume that $\sigma_A = G_m$ on all valid inputs.
\end{enumerate}
 
In agreement with our intuition described above, 
the following key lemma explicitly specifies $2^{3m}$ 
different inputs on which Bob and/or Carol have to send
different messages.  
Theorem~2 is immediate.

\beforethm
{\noindent \bf Lemma~1\quad}{\sl
Consider the above one-round protocol for computing~$G_m$.
Let $\sigma_B$ and $\sigma_C$ denote Bob's and Carol's messages
on inputs
${\mathbf y} = ({\mathbf y}_{\mathrm{high}},{\mathbf y}_{\mathrm{low}})$
and
${\mathbf z} = ({\mathbf z}_{\mathrm{high}},{\mathbf y}_{\mathrm{low}})$,
respectively.
Let $\sigma'_B$ and $\sigma'_C$ denote Bob's and Carol's messages
on inputs
${\mathbf y}' = ({\mathbf y}'_{\mathrm{high}},{\mathbf y}'_{\mathrm{low}})$
and
${\mathbf z}' = ({\mathbf z}'_{\mathrm{high}},{\mathbf y}'_{\mathrm{low}})$,
respectively.
Then the following holds.
\begin{enumerate}
\item[(i)] 
 If ${\mathbf y}_{\mathrm{high}} \neq {\mathbf y}'_{\mathrm{high}}$
      and 
      ${\mathbf y}_{\mathrm{low}} = {\mathbf y}'_{\mathrm{low}}$,
      then $\sigma_B \neq \sigma'_B$.
\item[(ii)] 
 If ${\mathbf z}_{\mathrm{high}} \neq {\mathbf z}'_{\mathrm{high}}$
      and 
      ${\mathbf y}_{\mathrm{low}} = {\mathbf y}'_{\mathrm{low}}$,
      then $\sigma_C \neq \sigma'_C$.
\item[(iii)]  
 If ${\mathbf y}_{\mathrm{low}} \neq {\mathbf y}'_{\mathrm{low}}$,
      then $\sigma_B \neq \sigma'_B$ or $\sigma_C \neq \sigma'_C$.
\end{enumerate} }
\afterthm

We first prove~(i) by contradiction.
Assume ${\mathbf y}_{\mathrm{high}} \neq {\mathbf y}'_{\mathrm{high}}$,
${\mathbf y}_{\mathrm{low}} = {\mathbf y}'_{\mathrm{low}}$,
and $\sigma_B = \sigma'_B$.
Let ${\mathbf x}$ be the unique input to Alice such 
that $G_m({\mathbf x}, {\mathbf y}, {\mathbf z}) = 1$.
Then
$({\mathbf x}, {\mathbf y}', {\mathbf z})  
  = {\mathbf (}{\mathbf x}, 
    ({\mathbf y}'_{\mathrm{high}}, {\mathbf y}_{\mathrm{low}}),
     {\mathbf z}{\mathbf )}$ 
is a valid input on which $G_m$ takes the value~0.
But, since $\sigma_B = \sigma'_B$, 
we also have $\sigma_C({\mathbf z},\sigma_B)
= \sigma'_C({\mathbf z}, \sigma'_B)$, and hence
Alice incorrectly outputs the same
answer $\sigma_A( {\mathbf x}, \sigma_B,\sigma_C)
 = \sigma_A( {\mathbf x}, \sigma'_B,\sigma'_C)$ in both cases.
Thus, the assumption is wrong and (i) holds.

The proof of~(ii) is almost identical to the proof of~(i), 
and we therefore omit~it.

We also prove~(iii) by contradiction.
Assume ${\mathbf y}_{\mathrm{low}} \neq {\mathbf y}'_{\mathrm{low}}$,
$\sigma_B = \sigma'_B$, and $\sigma_C = \sigma'_C$.
Let ${\mathbf x} = ({\mathbf x}_{\mathrm{high}},{\mathbf 0})$ 
be the unique input to Alice such that 
$G_m({\mathbf x}, {\mathbf y}, {\mathbf z}) = 1$.
Since the protocol correctly computes~$G_m$, then 
Alice must answer~1 on the 
input $({\mathbf x}, {\mathbf y}, {\mathbf z})$.
But then $({\mathbf x}, {\mathbf y}', {\mathbf z}')$ is also a valid
input on which Alice answers~1.
Further, let ${\mathbf x}' 
 = ({\mathbf x}'_{\mathrm{high}},{\mathbf x}'_{\mathrm{low}})$ 
be the unique input to Alice such that 
$G_m({\mathbf x}', {\mathbf y}', {\mathbf z}) = 1$.
Since the protocol correctly computes~$G_m$, then 
Alice must answer~1 on the 
input $({\mathbf x}', {\mathbf y}', {\mathbf z})$.
But then $({\mathbf x}', {\mathbf y}, {\mathbf z}')$ is also a valid
input on which Alice answers~1.

Thus, Alice answers~1 on all of these 4~valid inputs:
$({\mathbf x}, {\mathbf y}, {\mathbf z})$,
$({\mathbf x}, {\mathbf y}', {\mathbf z}')$,
$({\mathbf x}', {\mathbf y}', {\mathbf z})$, and
$({\mathbf x}', {\mathbf y}, {\mathbf z}')$.
But, since 
${\mathbf y}_{\mathrm{low}} \neq {\mathbf y}'_{\mathrm{low}}$,
then (as we show in the next paragraph) $G_m$ takes the value~0 
on at least one of the valid inputs
$({\mathbf x}, {\mathbf y}', {\mathbf z}')$ 
and $({\mathbf x}', {\mathbf y}, {\mathbf z}')$,
and thus the protocol incorrectly computes~$G_m$.
Hence, the assumption is wrong and (iii) follows.

To see that $G_m$ has to take the value~0 
on at least one of the valid inputs
$({\mathbf x}, {\mathbf y}', {\mathbf z}')$ 
and $({\mathbf x}', {\mathbf y}, {\mathbf z}')$,
assume otherwise.
Let $1 \leq i \leq m$ be a coordinate where 
${\mathbf y}_{\mathrm{low}}$ and ${\mathbf y}'_{\mathrm{low}}$
differ.
For ease of notation, we 
let $y_{\mathrm{low}}$ denote the $i$-th coordinate (bit) of 
${\mathbf y}_{\mathrm{low}} \in \{0,1\}^m$, 
and we use similar notation for the $i$-th coordinate of 
the other vectors.
\begin{itemize}
\item Since $G_m({\mathbf x}, {\mathbf y}, {\mathbf z}) =1$, then 
$(x_{\mathrm{high}} + y_{\mathrm{high}} +z_{\mathrm{high}} + y_{\mathrm{low}}) 
\textrm{ mod 2} = 0$.
\item Since $G_m({\mathbf x}', {\mathbf y}', {\mathbf z}) =1$, then 
$(x'_{\mathrm{high}} + y'_{\mathrm{high}} +z_{\mathrm{high}} + 1) 
\textrm{ mod 2} = 0$.
\item Since $G_m({\mathbf x}, {\mathbf y}', {\mathbf z}') =1$, then 
$(x_{\mathrm{high}} + y'_{\mathrm{high}} +z'_{\mathrm{high}} + 
   y'_{\mathrm{low}}) \textrm{ mod 2} = 0$.
\item Since $G_m({\mathbf x}', {\mathbf y}, {\mathbf z}') =1$, then 
$(x'_{\mathrm{high}} + y_{\mathrm{high}} +z'_{\mathrm{high}} + 1) 
  \textrm{ mod 2} = 0$.
\end{itemize}
But all of these 4~equations cannot hold at the same time,
and thus the assumption that $G_m$ takes the value~1
on $({\mathbf x}, {\mathbf y}', {\mathbf z}')$ 
and $({\mathbf x}', {\mathbf y}, {\mathbf z}')$ is wrong.
This completes our proof of Lemma~1, from which Theorem~2 
immediately follows.

It is worthy noticing that, by Lemma~1, 
for Alice to correctly output the value of~$G_m$, 
she has to be able to correctly compute $f$ on every one of the
$m$ instances of~$f$.  This is in general not so, and it is a
deep open question in communication complexity to characterize
the functions that possess this property.

\section{Multi- Rounds and Parties}\label{sec:manyparties}
We now generalize $f$ defined in Eq.~\ref{eq:modulofoursum} 
to a function~$F$ 
which we shall use to prove a separation in terms of 
the number of parties.
There are $k$ parties, where party $P_i$ obtains input data 
$x_i \in V= \{0,\dots,2^n-1\}$ ($i=1,\dots,k$).
We say that an input ${\bf x} = (x_1,\ldots,x_k)$ is {\em valid\/} if it
satisfies that  
\begin{equation}\label{eq:promise}
\left(\sum_{i=1}^k x_i\right) \textrm{mod $2^{n-1}$} = 0.
\end{equation} 
Let $F : V^k \rightarrow \{0,1\}$ denote the Boolean function 
on the valid inputs defined by 
\begin{equation} F({\bf x}) = \frac{1}{2^{n-1}}
\left[\left(\sum_{i=1}^k x_i\right) \textrm{mod $2^n$}\right]. 
\end{equation}
We say that a valid input {\bf x} is {\em $b$-valid\/} if $F({\bf x})=b$ 
($b=0,1$). 
The function $F$ can be viewed as computing the
$n$-th least significant bit of the sum of the~$x_i$'s.

We first show that with prior quantum entanglement, $k$ bits of
communication are necessary and sufficient for every party to 
evaluate~$F$.  That is, for all $k \geq 2$ and $n \geq 1$, 
\begin{equation}\label{eq:qcc}
Q(F,k,n)=k.
\end{equation}
Then, we show how the parties can evaluate $F$ with roughly $k \log_2 k$ 
bits of communication without using any entangled particles.
Specifically, for all $k \geq 2$ and $n \geq 1$, 
\begin{equation}\label{eq:ccc-upper}
C(F,k,n) \leq (k-1) \{\lceil \log_2(k-1) \rceil + 1\} +1.
\end{equation}
Finally, we prove that this is optimal up to low order terms
by showing that, for all $k \geq 2$ and $n \geq \log_2 k$, 
\begin{equation}\label{eq:ccc-lower}
C(F,k,n) > k\log_2(k) -k.
\end{equation}
By comparing the bounds of Eqs.~\ref{eq:qcc} and~\ref{eq:ccc-lower},
we see that we have established a separation by a factor 
of~$\log_2(k/2)$.

\subsection{With Entanglement}\label{subsec:withentanglement}

We first show that if the parties share entangled particles, 
then in a straight-forward manner, the $k$ parties can evaluate $F$
using only one bit of communication each.  This is obtained by a direct
generalization of the idea used both in Sect.~2.1 of~\cite{BCD97}
(which itself is based on the work of Mermin~\cite{Mermin90})
and in~\cite{Grover97}.
The prior quantum entanglement shared by the $k$ parties is the cat
state 
$\ket{q_1 \ldots q_k} = (\ket{0 \ldots 0} + \ket{1 \ldots 1})/\sqrt{2}$,
where party $P_i$ holds qubit $q_i$ ($i=1,\ldots,k$).

Each party $P_i$ uses the following procedure.
First party $P_i$ applies a phase-change operator 
$\phi(x_i)$ defined by $\ket{0} \mapsto \ket{0}$ and 
$\ket{1} \mapsto \exp(2 \pi x_i \sqrt{-1} / 2^n) \ket{1}$
on her qubit~$q_i$.
Thanks to the promise on the inputs, 
these phase rotations add up so that the resulting state is
$(\ket{0 \dots 0} + (-1)^{F({\bf x})} \ket{1 \dots 1})/\sqrt{2}$.
Then she applies the Walsh-Hadamard transform that maps 
\ket{0} to $(\ket{0} + \ket{1})/ \sqrt{2}$, and
\ket{1} to $(\ket{0} - \ket{1})/ \sqrt{2}$.
Finally, she measures her qubit $q_i$ in the computational basis 
$\{\ket{0},\ket{1}\}$ and broadcasts the outcoming bit.

Let $b_i$ be the outcome of party $P_i$'s measurement.
Simple calculations  show that $b_1 \oplus \cdots \oplus b_k$
equals $F(x_1,\ldots,x_k)$, where $\oplus$ denotes addition in modulo-2
arithmetic.  
It~follows that every party can compute the value of $F$ from
the $k$ communicated bits.  On the other hand, $k$ bits of communication
are necessary since if one of the parties does not broadcast any bits,
then none of the others can determine the value of~$F$.  To see this,
note that if we toggle the most significant bit of any one of the inputs,
then the value of $F$ changes.  
Equation~\ref{eq:qcc} follows.

\subsection{Without Entanglement}

The simplest way to evaluate the function $F$ is for all but one of
the parties to broadcast their inputs.  The last party then 
evaluates $F(x_1,\ldots,x_k)$ and communicates the resulting bit to the
others.  Hence, the communication complexity (without entanglement) is
at most $(k-1) n +1$.

Now, consider that all but one of the parties broadcast the $d$
most significant bits of their inputs, for some integer $d \geq 1$.  
The last party, say $P_k$, then computes the sum 
$\big(\sum_{i=1}^k x_i\big) -\delta$ where 
\[\delta = \sum_{i=1}^{k-1} (x_i \textrm{ mod $2^{n-d}$}).\]
Suppose $n \geq d$ where $d = 1 + \lceil \log_2 (k-1) \rceil$.
Then
\[0 \leq \delta \leq (k-1)(2^{n-d}-1) < 2^{n-1},\]
so party $P_k$ knows the value of the sum $\sum_{i=1}^k x_i$ up to
an additional non-negative term strictly smaller than $2^{n-1}$.  Since
the sum is divisible by $2^{n-1}$ for all valid inputs, 
party $P_k$ can determine it exactly and thus compute the value of~$F$. 
It follows that $(k-1) d +1$ bits of communication suffice, as stated 
as Eq.~\ref{eq:ccc-upper}.

A good method to prove lower bounds for the communication complexity of 
functions comes from a combinatorial view on the protocol for the
communication.  Consider the space $V^k$ of all possible inputs,
where $V = \{0,\ldots,2^n-1\}$.
A~{\em rectangle\/} in $V^k$ is a subset $R \subseteq V^k$ such that
$R = R_1 \times \cdots \times R_k$ for some $R_i \subseteq V$
($i=1,\ldots,k$). 
If~a rectangle contains no 0-valid inputs or no 1-valid 
inputs, then it is said to be $F$-{\em monochromatic}.

We now use the observation that every deterministic and 
errorless communication
protocol corresponds to a covering of all the valid inputs in 
$V^k$ by $F$-monochromatic rectangles (see~\cite{KN97}).
Without increasing the communication complexity, such a protocol 
can always be transformed into a protocol that uses a partitioning 
that covers all of $V^k$, and for which each monochromatic rectangle
 contains at least one valid input. 
By proving that every such partition requires at least $t$ 
rectangles, we also prove that the communication 
complexity of $F$ is at least $\log_2 t$~\cite{KN97}.
Hence, upper bounds on the cardinality of the possible $F$-monochromatic
rectangles imply a lower bound on the communication complexity of~$F$.  

In the appendix, we prove that if a rectangle $R \subseteq V^k$ 
is $F$-monochromatic and if $R$ contains a valid input,
then its cardinality is upper bounded by a value $r$, for which 
\begin{equation}\label{eq:cardinality}
r = \left(\frac{2^n - 2}{k} +1\right)^k.
\end{equation}
Since there are $2^{nk}$ input values to be covered, 
this bound on the size of the rectangles shows that we 
need at least $t=2^{nk}/r$ rectangles to partition 
$V^k$ in the above described fashion.

If $n \geq \log_2 k$ and $k \geq 2$, then basic algebra gives that  
\[\log_2 t = \log_2 \left(\frac{2^{nk}}{r}\right) 
  > k\log_2(k)-k.\] 
{From} this, the lower bound on the communication complexity 
of Eq.~\ref{eq:ccc-lower} follows.

\section*{Acknowledgements}
We are grateful to Lucien Hardy for 
pointing out an error in an earlier version of this paper,
and to Richard Cleve and Andrew Landahl for discussions. 
H.\,B.{} thanks Ricard Gavald\`a for stimulating conversations.

W.\,v.D. has been supported by the European TMR Research Network 
EPR-4061PL95-1412,
Hewlett-Packard, and the Institute for Logic, Language, and Computation 
in Amsterdam.
P.\,H.{} has been supported in part by the {\sc esprit} Long
Term Research Programme of the EU under project number 20244 
({\sc alcom-it}).  
A.\,T.{} has been supported in part by a postgraduate fellowship from
Canada's {\sc nserc}.
Parts of this research were carried out while P.\,H.{} 
was at the Laboratoire d'informatique th{\'e}orique et quantique at
Universit{\'e} de Montr{\'e}al, and during the 1997
Elsag-Bailey\,--\,I.S.I.{} Foundation research meeting on quantum
computation.


\section*{Appendix: Upper Bound on the Cardinality of a Mono\-chromatic 
Rectangle}
Equip the set $V =\{0,\ldots,2^n-1\}$ 
with the natural addition operation, denoted $\oplus$
and given by $x \oplus y = (x + y) \textrm{ mod $2^n$}$.
Then $V = \langle V,\oplus \rangle$ is a cyclic group of order~$2^n$. 

Let $R \subseteq V^k$ be a fixed rectangle.  By definition, 
$R = R_1 \times \cdots \times R_k$ for some subsets $R_i \subseteq V$, 
$i=1,\ldots,k$.  
For any two subsets $A,B \subseteq V$, define 
$A \oplus B = \{a \oplus b \mid a \in A, b \in B\}$.
We now define a family of subsets of~$V$.
Set $S_0 = \{0\} \subset V$ 
and $S_i = S_{i-1} \oplus R_i$ for $i=1,\ldots,k$.
Then for each element $(x_1,\ldots,x_k) \in R$, the value
$\big(\sum_{i=1}^k x_i\big) \textrm{ mod $2^n$}$ is in~$S_k$.
We shall use Kneser's theorem~\cite{KneserMann} to
give an upper bound on the cardinality of~$R$.

\beforethm
{\noindent \bf Kneser's theorem\quad}{\sl
Let $G = \langle G, \oplus \rangle$ be an Abelian group with finite 
subsets $A$ and~$B$.  Then there exists a subgroup $H$ of $G$ such that
\[A \oplus B \oplus H = A \oplus B\]
and 
\[|A \oplus B| \geq |A \oplus H| + |B \oplus H| - |H|.\]}
\afterthm

Let $H_i$ be the largest subgroup of $V$ for which
$S_i = S_i \oplus H_i$, ($i = 0,\ldots,k$).  
Since $\oplus$ is associative, then
$H_{i-1} \subseteq H_i$ for all $1 \leq i \leq k$. 

Suppose $R$ is a monochromatic rectangle that contains a valid input.
Without loss of generality, assume that it is a 0-valid input,
that is, that $0 \in S_k$.  Then $H_i$ is the trivial subgroup 
$\{0\}$ for all $i$, 
since otherwise we have that $2^{n-1} \in H_i \subseteq H_k$ and hence 
$R$ would not be monochromatic. 
This shows that if we identify $A=S_{i-1}$ and $B=R_i$ in Kneser's
theorem, it follows that $H$ is the trivial subgroup.
We therefore have that
$|S_i| \geq |S_{i-1}| + |R_i| -1$, so
\[|S_k| \geq \sum_{i=1}^k |R_i| - (k-1).\]
Since $2^{n-1} \not\in S_k$, then $|S_k| \leq 2^n -1$, so
\[\sum_{i=1}^k |R_i| \leq 2^n - 2 + k,\]
and therefore
\[|R| = \prod_{i=1}^k |R_i| \leq \left(\frac{2^n - 2}{k} +1\right)^k.\] 
It follows that the right hand side of Eq.~\ref{eq:cardinality} is 
an upper bound on the cardinality of any $F$-monochromatic rectangle 
that contains a valid input.


\end{document}